# Making a Wavefunctional representation of physical states congruent with the false vacuum hypothesis of Sidney Coleman

A. W. Beckwith

Department of Physics and Texas Center for Superconductivity and Advanced Materials at the University of Houston
Houston, Texas 77204-5005 USA

**ABSTRACT**

We examine quantum decay of the false vacuum in the driven sine-Gordon system. Using both together permits construction of a Gaussian wave functional; this is due to changing a least-action integral with respect to the WKB approximation. In addition, we find that after rescaling, the soliton-antisoliton (S-S') separation distance obtained from the Bogomol'nyi inequality permits a dominant $\phi^2$ contribution to the least action integrand. This is from an initial scalar potential characterized by a tilted double well potential construction.

Correspondence: A. W. Beckwith:    **projectbeckwith2@yahoo.com**

# INTRODUCTION

We apply the vanishing contribution to a physical system of a topological charge Q to show how the Bogomol'nyi inequality[1,2] can be used to simplify a Lagrangian potential energy term. This is so that the potential energy is proportional to a quadratic $\phi^2$ scalar field contribution. In doing so, we work with a field theory featuring a Lorenz scalar singlet valued field in D+1 dimensional spacetime. Here, the D is the spatial dimensions of the analyzed system, so if D = 1 we are working with 1 spatial dimension plus a time contribution.

We then describe contribution of the quantum decay of a false vacuum.[1,3] For forming the Gaussian wavefunctionals in our new functional integration presentation of our generalized rate creation problem, we employed a least action principle that Sidney Coleman[3] used for WKB-style modeling of tunneling.

As a sign of its broad scientific interest, for over two decades several quantum tunneling approaches[3] have been proposed to this issue of the quantum decay of the false vacuum. One[3] is to use functional integrals to compute the Euclidean action ("bounce") in imaginary time. This permits inverting the potential and modifying what was previously a potential barrier separating the false and true vacuums into a potential well in Euclidean space and imaginary time. J.H. Miller Jr.[4] applied another approach, using the Schwinger[5] proper-time method, to calculate the rates of particle-antiparticle pair creation in an electric field for simplifying transport problems. We also mathematically elaborate upon the S-S' domain wall paradigm[6,7] so that a rate equation formalism, which uses this Gaussian wavefunctional derivation, has the kinetic energy information for our rate of transfer problem. The derived wavefunctionals contain the tilted potential contribution the false vacuum hypothesis; this gives us physics problems that are re scaled to a quadratic $\phi^2$ scalar field contribution.

## BASIC TECHNIQUES

We apply the domain wall physics of S-S' pairs to obtain a quadratic scalar valued potential for transport physics problems involving weakly coupled scalar fields. When we modify the energy/mass representation of the soliton kink by the Bogomol'nyi inequality,[1] we can use the bound on our modified potential to simplify a Euclidian least action integral.

When we use Euclidian imaginary time,

$$\int D\phi \cdot \exp\left( (i/\hbar) \cdot \int d^d x \cdot \left[ \frac{1}{2} \cdot (\partial \phi)^2 - V(\phi) \right] \right) \rightarrow \qquad (1a)$$

transforms to

$$\int D\phi \cdot \exp\left( (-1/\hbar) \cdot \int d_E^d x \cdot \left[ \frac{1}{2} \cdot (\partial \phi)^2 + V(\phi) \right] \right) \qquad (1b)$$

We should note that Eq. (1b) has an energy expression of the form

$$\varepsilon(\phi) \equiv \int d_E^d x \cdot \left[ \frac{1}{2} \cdot (\partial \phi)^2 + V(\phi) \right] \qquad (2a)$$

Eq. (2a) has a potential term that we can write as

$$V(\phi) \equiv C_0 \cdot (\phi - \phi_0)^2 + C_1 \cdot (\phi - \phi_0)^4 + H.O.T. \qquad (2b)$$

Furthermore, even when we invert our potentials, we can simplify our expression for the potential by procedures that eliminate the scalar potential terms higher than $\phi^2$ by considering the energy per unit length of a soliton kink. After rescaling to different constants, this is given by A. Zee[2] as

$$\tilde{\varepsilon}(x) = \frac{1}{2} \cdot \left(\frac{d \cdot \phi}{d \cdot x}\right)^2 + \frac{\lambda}{4} \cdot \left(\phi^2 - \varphi\right)^2 \tag{3}$$

with a mass of the kink or antikink given by

$$M \equiv \int dx \cdot \tilde{\varepsilon}(x) \tag{3a}$$

to be bounded below, namely, by use of the Bogomol'nyi inequality

$$M \geq \int dx \cdot \sqrt{\frac{\lambda}{2}} \cdot \left|\left(\frac{d \cdot \phi}{d \cdot x}\right) \cdot \left(\phi^2 - \varphi^2\right)\right| \geq \left|\frac{4}{3 \cdot \sqrt{2}} \cdot \mu \cdot \left(\frac{\mu^2}{\lambda}\right) \cdot Q\right| \tag{4}$$

where Q is a topological charge of the domain wall problem and $\mu \propto \sqrt{\lambda \cdot \phi_0^2}$. The $\lambda$ relates to the dimensional magnitude of a double-well potential; we are assuming a small tilt to this double-well potential problem due to an applied electric field. We define conditions for forming a wave functional by Eq. (5):[1,8]

$$\Psi \equiv c \cdot \exp(-\alpha \cdot \int dx^{(D \equiv 1)} [\phi_0 - \phi_C]^2) \tag{5}$$

We presuppose, when we obtain Eq. (5), a power series expansion of the Euclidian Lagrangian, $L_E$ about $\phi_0$. Here, $\phi_C$ is giving us local minimum values for the physical system with respect to minimum values of our potential, with or without an applied electric field tilting the double well potential. The first term of this expansion,

$$L_E \big|_{\phi=\phi_O} = \frac{1}{2} \cdot (\vec{\nabla}\phi)^2 \big|_{\phi=\phi_0} \equiv \varepsilon(\phi) \big|_{\phi \equiv \phi_0} \tag{6}$$

is a comparatively small quantity that we may ignore most of the time. Furthermore, we simplify working with the least-action integral by assuming an almost instantaneous nucleation of the S-S' pair. We may then write, starting with a Lagrangian density $\zeta$,

$$\int d\tau \cdot dx \cdot \zeta \to t_P \cdot \int dx \cdot \zeta \to t_P \cdot \int dx \cdot L \tag{7}$$

Quantity $t_P$ in Eq. (7) is scaled to unity. Eq. (7) allows us to write our wave functional as a one-dimensional integrand; in this calculation, $t_P \propto 1$ is a time-unit interval.

Eq. (7) needs considerable explanation. To do this, break up the Lagrangian density as

$$\zeta \equiv (\partial \phi)^2 + V(\phi) \tag{8}$$

with (for a very brief instant of time)

$$(\partial \phi)^2 = (\partial \phi / \partial \tau_E)^2 + (\vec{\nabla} \phi)^2 \tag{9}$$

becomes

$$(\partial \phi)^2 \cong + (\vec{\nabla} \phi)^2 \tag{9a}$$

Look at the integrand as

$$\int d\tau \cdot dx \cdot \zeta = t_P \cdot \breve{\varepsilon}(\phi) \tag{9b}$$

with

$$\breve{\varepsilon}(\phi) \equiv \int dx \cdot \left[ \frac{1}{2} \cdot (\nabla \phi)^2 + V(\phi) \right] \tag{10}$$

Then we have to look at the behavior of

$$(\nabla \phi) \equiv \delta_n(x - L/2) - \delta_n(x + L/2) \tag{11}$$

which would represent the behavior of test functions converging to Dirac delta functions as $n \to \infty$. Furthermore, if N is very large

$$\int dx \cdot \left[\frac{1}{2} \cdot (\nabla \phi)^2\right] \xrightarrow[n \to N \approx \infty]{} \frac{1}{2} \cdot \int dx \cdot [\delta_N(x - L/2) - \delta_N(x + L/2)]^2$$
$$\cong \frac{1}{2} \cdot [2/\sqrt{2}] \equiv \frac{1}{\sqrt{2}} < 1 \tag{12}$$

when for all N we are writing

$$\delta_N(x \pm L/2) \equiv \delta_N(\tilde{x}) \equiv (N/2 \cdot \sqrt{\pi}) \cdot \exp(-\tilde{x}^2 \cdot N^2/4) \tag{13}$$

where for all $N$ values we have

$$\int_{-\infty}^{+\infty} \delta_N(\tilde{x}) \cdot d\tilde{x} = 1 \tag{14}$$

So, then, we are analyzing this problem with contributions about the domain walls of $x \cong \pm L/2$ assuming a thin wall approximation, as illustrated by Fig. 1.

**[Insert Fig. 1 here]**

Introducing domain wall physics via Eq. (6) and Eq. (7) allows us to write our wave functional as proportional to[1,8]

$$\psi \propto c \cdot \exp\left(-\tilde{\beta} \cdot \int L \, d\tau\right) \tag{15}$$

with the Lagrangian treated as a typical Taylor series expansion w.r.t. the $\phi_0$ input scalar value

$$L_E \approx L_E \mid_{\phi = \phi_0} + \frac{1}{2} \cdot (\phi - \phi_0)^2 \frac{\partial^2 \cdot V_E}{\partial \cdot \phi^2} \mid_{\phi = \phi_0} + \frac{1}{3!}(\phi - \phi_0)^3 \cdot \frac{\partial^3 \cdot V_E}{\partial \cdot \phi^3} \mid_{\phi = \phi_0} +$$
$$\frac{1}{4!}(\phi - \phi_0)^4 \cdot \frac{\partial^4 \cdot V_E}{\partial \cdot \phi^4} \mid_{\phi = \phi_0} \tag{16}$$

We should be aware that for a wick rotation, when $t = -i \cdot \tau_E$ that for $d$ dimensions $d^d x = -i \cdot d_E^d x$ with $d_E^d x = d\tau_E \cdot d^{d-1} x$. We also use a conserved current quantity of[1,9]

$$J^{\mu} = \frac{1}{2 \cdot \varphi} \cdot \varepsilon^{\mu\nu} \cdot \partial_{\nu} \cdot \phi \tag{17}$$

with a *topological* charge of[1]

$$Q \equiv \int_{-\infty}^{+\infty} dx \cdot J^{0}(x) = \frac{1}{2 \cdot \varphi} [\phi \cdot (\infty) - \phi \cdot (-\infty)] \tag{18}$$

where our contribution of topological charge from $Q \to \varepsilon^{+} \approx 0$ means mesonic behavior

In Zee,[2] the $\varphi$ term is due to his setting of two minimum positions for $\phi$ for a double well potential. It is useful to note that if we look at the mass of a kink via a scaling $\mu \propto \sqrt{\lambda \cdot \phi_0^2}$ with M defined as the same as the energy of a soliton kink given in Eq. (3), with a subsequent mass given in Eq. (3a), that we have, via using $a^2 + b^2 \geq 2 \cdot |a \cdot b|$, an inequality of the form given by Eq. (4). So that[2]

$$M \geq |Q| \tag{19}$$

with *mass* M in terms of units of $\frac{4}{3 \cdot \sqrt{2}} \cdot \mu \cdot \left(\frac{\mu^2}{\lambda}\right)$. If we note that we have

$(\phi - \phi_0)^4 = (\phi^2 - \phi_0^2)^2 - 4 \cdot \phi \cdot \phi_0 \cdot (\phi - \phi_0)^2$ in one dimension, we physically use our topological current as a very small quantity. Then, we can write

$$L_E \geq |Q| + \frac{1}{2} \cdot (\phi_0 - \phi_C)^2 \cdot \{\ \} \tag{20}$$

where

$$|Q| \to \varepsilon^{+} \geq 0 \tag{21}$$

Due to a *topological* current argument (S-S' pairs usually being of opposite charge) and

$$\{\ \} \equiv \{\ \}_A - \{\ \}_B \equiv 2 \cdot \Delta E_{gap} \approx 2 \cdot \alpha^{-1} \tag{22}$$

where if we pick[1]

$$\frac{(\{\ \} \equiv \{\ \}_A - \{\ \}_B)}{2} \equiv \Delta E_{gap} \equiv V_E(\phi_F) - V_E(\phi_T) \tag{23}$$

This means a wavefunctional with information from a inverted potential as part of a transport problem of weakly coupled systems along the lines suggested by Ciraci and Tekeman.[10] We found for $D = 1$, can write more generally the initial configuration of the form[11]

$$\Psi_i[\phi(\mathbf{x})]\big|_{\phi \equiv \phi_{Ci}} = c_i \cdot \exp\{-\alpha \int d\mathbf{x} [\phi_{ci}(\mathbf{x}) - \phi_0]^2\}, \tag{24a}$$

which is

$$\Psi_i[\phi(\mathbf{x})] = c \cdot \exp\{-\alpha' \int d\mathbf{x}\, L_E(\mathbf{x})\}$$
$$= c \cdot \exp\{-\alpha' S_E\}. \tag{24b}$$

in addition, we would also have a final state immediately after tunneling,[1,11]

$$\Psi_f[\phi(\mathbf{x})]\big|_{\phi \equiv \phi_{Cf}} = c_f \cdot \exp\{-\int d\mathbf{x}\, \beta(\mathbf{x})[\phi_{Cf}(\mathbf{x}) - \phi_0(\mathbf{x})]^2\}, \tag{24c}$$

Furthermore, we have that[1] in the case of a driven sine-Gordon potential a situation in which we can approximate the physics of Fig. 1 by

$$\frac{\partial V}{\partial \phi} = 0, \tag{25a}$$

$$\Rightarrow \phi_F \approx \left[\frac{\varepsilon^+}{\varepsilon^+ + 1}\right] \approx \varepsilon^+$$

when we are working with , when[8] (assuming $C_a \gg C_b$)

$$V(\phi) \approx C_a \cdot (1 - \cos\phi) + C_b \cdot (\phi - \phi_{Ci,f})^2 \tag{25b}$$

We also have in the case of a driven sine-Gordon potential a situation where we can generalize our wavefunctionals as[1,11]

$$\Psi_i[\phi(\mathbf{x})]\Big|_{\phi \equiv \phi_{Ci}} = c_i \cdot \exp\left\{-\alpha \int d\mathbf{x} [\phi_{ci}(\mathbf{x}) - \phi_0]^2\right\} \to$$
$$c_1 \cdot \exp\left(-\alpha_1 \cdot \int d\tilde{x} [\phi_F]^2\right) \equiv \Psi_{initial}, \tag{26a}$$

and

$$\Psi_f[\phi(\mathbf{x})]\Big|_{\phi \equiv \phi_{Cf}} = c_f \cdot \exp\left\{-\int d\mathbf{x}\, \alpha \left[\phi_{cf}(\mathbf{x}) - \phi_0(\mathbf{x})\right]^2\right\} \to$$
$$c_2 \cdot \exp\left(-\alpha_2 \cdot \int d\tilde{x} [\phi_T]^2\right) \equiv \Psi_{final}, \tag{26b}$$

where a driven sine-Gordon system is of the form [9] (assuming $C_a \gg C_b$)

$$= \phi_T \approx 2 \cdot \pi \tag{27}$$

## CONCLUSION

It is straightforward to construct wavefunctionals that represent creation of a particular event within an embedding space. Diaz and Lemos[12] use this technique as an example of the exponential of a Euclidian action to show how black holes nucleate from nothing. This was done in the context of de Sitter space. This trick was also used by Kazumi Maki[13] to observe a field theoretic integration of condensates of S-S' pairs in the context of boundary energy of a two-dimensional bubble of space-time. This two-dimensional bubble action value was minus a contribution to the action due to volume energy of the same two-dimensional bubble of space-time. Maki's[13] probability expression for S-S' pair production is not materially different from what Diaz and Lemos[12] used for black hole nucleation.

What we have done is to generalize this technique to constructing wavefunctional representations of false and true vacuum states in a manner that allows for transport problems to be written in terms of kinetic dynamics as they are given by a functional generalization of a tunneling Hamiltonian. It also allows us to isolate soliton/instanton information in a potential field that overlaps with a Gaussian wavefunctional presentation of soliton/instanton dynamics.[1,2] We believe that this approach will prove especially fruitful when we analyze nucleation of instanton[14] states that contribute to lower dimensional analysis of the configurations of known physical systems (e.g., $NbSe_3$).[1,2,8] This approach to wavefunctionals materially contributes to calculations we have performed with respect to I-E curves fitting experimental data quite exactly[1,2,8] — and in a manner not seen in more traditional renderings of transport problems in condensed matter systems with many weakly coupled fields interacting with each other.[15]

# FIGURE CAPTION

**Fig. 1:** Evolution from an initial state $\Psi_i[\phi]$ to a final state $\Psi_f[\phi]$ for a double-well potential (inset) in a 1-D model, showing a kink-antikink pair bounding the nucleated bubble of true vacuum. The shading illustrates quantum fluctuations about the initial and final optimum configurations of the field; $\phi_0(x)$ represents an intermediate field configuration inside the tunnel barrier. The upper right hand side shows how the fate of the false vacuum hypothesis gives a difference in energy between false- and true-potential vacuum values which we tie in with the results of the Bogomol'nyi inequality.

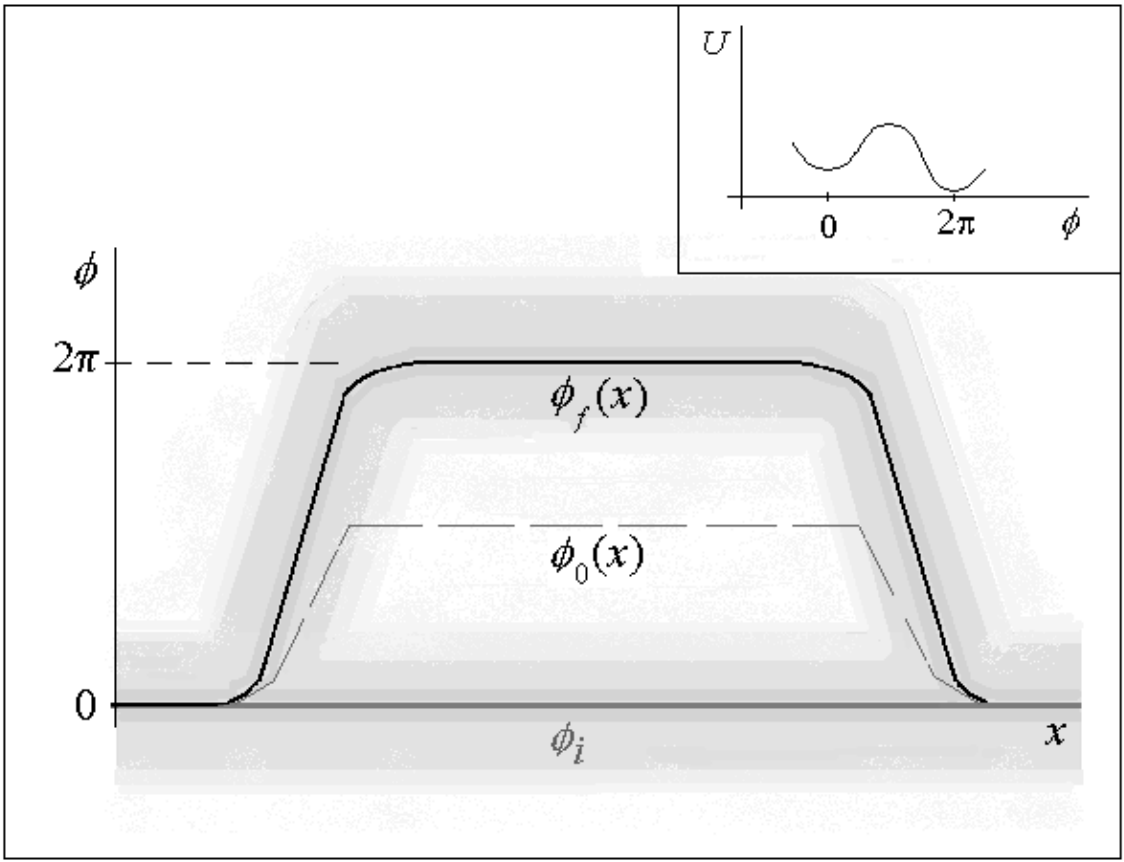

**Figure 1**

**Beckwith**